\documentclass[prd,twocolumn]{revtex4}
\usepackage{graphics}
\usepackage{bm}

\begin{document}

\title{
Eternal inflation and energy conditions in 
de Sitter spacetime\footnote{
Based on Ref.~\cite{GutVacWin??}} 
}

\author{Tanmay Vachaspati}
\affiliation{
Center for Education and Research in Cosmology and Astrophysics,
Department of Physics, Case Western Reserve University,
10900 Euclid Avenue, Cleveland, OH 44106-7079, USA.}


\begin{abstract}
Eternal inflation is shown to require violations of the Null 
Energy Condition (NEC) on superhorizon scales. With light scalar 
fields as the matter sources in a de Sitter background, there can 
be no classical or semiclassical violations of the NEC. However,
quantum fluctuations of the energy-momentum tensor, described by 
an expectation value of four field operators, do lead to large-scale 
NEC violations. The backreaction of such quantum fluctuations on 
the inflating spacetime is generally deduced at a heuristic level 
and leads to the eternal inflation scenario. A rigorous treatment
of the backreaction will necessarily include fluctuations of the
metric, and several new effects are expected to come into play.
\end{abstract}


\maketitle

\section{Eternal inflation scenario}
\label{scenario}

Inflation is based on the dynamics of a scalar field in an 
expanding universe. For example, as shown in Fig. 1, 
the homogeneous mode of a massive, non-interacting, scalar 
field, $\phi$, could be high up on the potential, $V(\phi )$,
at some initial epoch. Assuming that the dynamics is dominated
by the homogeneous mode, the large potential energy of the field 
would cause exponential expansion (inflation) as long as the 
rate at which the field rolls down the potential is slow. As 
the field rolls, the potential energy diminishes, and the Hubble 
expansion slows down.

\begin{figure}
\scalebox{0.60}{\includegraphics{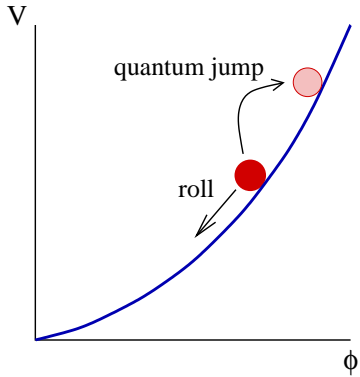}}
\caption{\label{eternalidea} 
Illustration of the inflationary and eternal inflationary
scenarios.}
\end{figure}

Eternal inflation modifies this picture of inflation by
acknowledging that the field is really undergoing quantum
dynamics. This means that there are uncertainties associated
with the location of the field. Occasionally the field need
not roll down the potential; instead it can jump {\em up}
the potential. After this jump, the field is higher up,
the potential energy is greater, and if the kinetic energy
is not too large, the Hubble expansion speeds up.
This is the basic idea of eternal inflation 
\cite{Ste83,Vil83,Lin86a,Lin86b,Lin86c,Lin87,GonLin87,GonLinMuk87}.

Developing the scenario further, we can expect the same
sort of behavior from scalar field modes that are smooth
on large scales but not completely homogeneous. Then there
will be regions in which the field will jump up. Such regions
will form bubbles of faster inflation in a background of
slower inflation. Indeed, even as the field in some regions
rolls down to the minimum value of the potential, there will 
always be bubbling regions, and inflation will be eternal. 
The universe as we see it will only develop in regions that 
manage to roll down all the way and thermalize.

\section{Why bother with eternal inflation?}
\label{why}

The eternal inflation scenario is attractive from a number
of viewpoints. To the particle physicist or string theorist
who is interested in cosmology, eternal inflation is simply 
a consequence of having a scalar field in a de Sitter background. 
If this simple combination results in eternal inflation, then 
it is very compelling and of fundamental importance. To the 
cosmologist, eternal inflation relieves certain concerns about 
the initial conditions that are essential for inflation 
\cite{Pir86,KunBra89,KunBra90,GolPir90,Gol91,VacTro98,TroVac99,
Vac99,BerGor01,LueStaVac03}. No matter how the universe started 
out, once it undergoes even a bit of inflation, eternal inflation 
sets in.  So the initial conditions become irrelevant \cite{Gut00}. 
To the general physicist,
eternal inflation leads to a radically new picture of the
universe, one which is bubbling forever, where new universes
are constantly forming. This is the ``multiverse'' picture.
Finally, to the observer, under certain circumstances and
with certain assumptions, eternal inflation predicts a
distribution of cosmological and other parameters that
can be measured \cite{GarVil03,TegVil03}. 

\section{The null energy condition}
\label{nec}

In a Friedman-Robertson-Walker universe, the equation
governing the Hubble expansion, $H$, is:
\begin{equation}
{\dot H} = -4\pi G (\rho + p ) + \frac{k}{a^2}
\label{dotH}
\end{equation}
where all the symbols have their usual meanings.
In an inflating spacetime, the curvature term can be
ignored and so we drop the last term. Then
\begin{equation}
{\dot H} > 0 \Rightarrow \rho + p < 0
\label{dotH>0}
\end{equation}
We can write the latter condition in terms of the
energy-momentum tensor $T_{\mu\nu}$ contracted with any
null vector, $N^\mu$:
\begin{equation}
N^\mu N^\nu T_{\mu\nu} < 0
\label{neccondition}
\end{equation}
In other words, the null energy condition (NEC) --
$N^\mu N^\nu T_{\mu\nu} \ge 0$ -- has to be violated if 
eternal inflation (${\dot H} > 0$) is to happen \cite{BorVil97}.

\begin{figure}
\scalebox{0.60}{\includegraphics{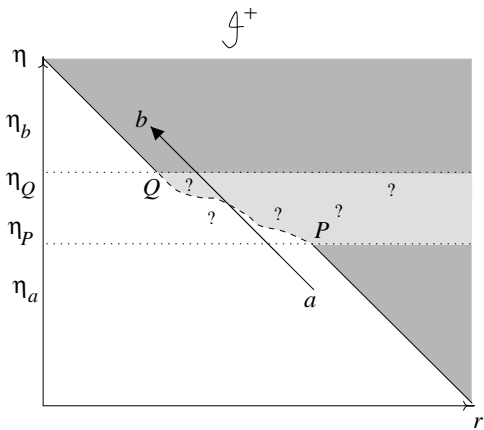}}
\caption{\label{eternalspacetime} Spacetime diagram for eternal 
inflation. $\eta$ is the conformal time and $r$ is the
comoving radial coordinate. Angular coordinates have
been suppressed. At early times the solid 
line shows the radius of the minimal antitrapped sphere
(MAS) in de Sitter space. The dashed portion of the line 
and question marks between the points \protect$P\protect$ 
and \protect$Q\protect$ represents the unknown position 
of the MAS during an upward jump. After the jump, the
inverse horizon decreases but the spacetime is again
de Sitter. So the MAS is again described by the solid
line but the intercept on the $r=0$ axis is smaller
(since $H^{-1}$ is smaller). A bundle of null geodesics
traveling from point $a$ to $b$ goes from being 
converging at $a$ to diverging at $b$, in violation
of the NEC.}
\end{figure}

The conclusion that one needs NEC violations to get eternal
inflation can also be deduced from a spacetime diagram relevant 
to eternal inflation (see Fig.~\ref{eternalspacetime}).

\section{Source of NEC violations?}
\label{source}

We have seen that eternal inflation needs NEC violations.
Now we try and determine if there exists a source for such 
violations. We will work within the context of a scalar field 
theory that is minimally coupled to gravity. The action is:
\begin{equation}
S = \int d^4x \sqrt{-g} 
   \biggl [ -\frac{1}{16\pi G} R + 
         \frac{1}{2} (\partial_\mu \phi )^2 - \frac{m^2}{2} \phi^2 
   \biggr ]
\label{action}
\end{equation}
The scalar field, $\phi$, has energy-momentum tensor
\begin{equation}
T_{\mu\nu} = \nabla_\mu \phi \nabla_\nu \phi - 
              \frac{1}{2} g_{\mu\nu} \left (
               \nabla^\mu \phi \nabla_\mu \phi - m^2 \phi^2
                                      \right ) \ .
\label{Tmunu}
\end{equation}

\subsection{Classical field}
\label{classical}

If $N^\mu$ is any null vector ($g_{\mu\nu}N^\mu N^\nu =0$)
then
\begin{equation}
N^\mu N^\nu T_{\mu\nu} = (N^\mu \partial_\mu \phi )^2 \ge 0
\label{NNTclassical}
\end{equation}
and hence a classical scalar field cannot violate the
NEC.

So, not surprisingly, to get eternal inflation, we must consider 
quantum field theory.

\subsection{Semiclassical gravity}
\label{semiclassical}

In quantum field theory, the energy-momentum tensor becomes an
operator. In semiclassical gravity, however, the metric is still 
classical. To relate the classical metric, to the quantum 
energy-momentum tensor, we use the semiclassical version of
Einstein's equation:
\begin{equation}
G_{\mu \nu} = 8\pi G \langle s| T_{\mu \nu} |s \rangle
\label{scEeqn}
\end{equation}
There are a few points about this equation that need to be
explained. First, it is essential to specify the state, $|s\rangle$,
of the quantum fields. In our case, we will take it to be
the Bunch-Davies vacuum, as is normally done in inflationary
calculations \cite{BraHil86,Albetal94}, and denote this state as 
$|0\rangle$ (It is well-known that with a non-vacuum choice 
of the state, the NEC can be violated \cite{EpsGlaJaf65,MorTho88,Kuo97} 
though the violations must still satisfy certain inequalities 
\cite{For93,ForRom95}.) 
Second, the expectation value of the energy-momentum
tensor is going to be infinite. Hence we need to carry out 
a suitable renormalization procedure. For example, we will
have to introduce a bare cosmological constant term that can 
be used to cancel out the infinite vacuum energy. The gravitational
coupling constant will also be normalized. Once all these 
necessary renormalization procedures are carried out, we may
write:
\begin{equation}
G_{\mu \nu} = 8\pi G_N \langle 0| T_{\mu \nu}^{ren} |0 \rangle
\label{scEeqnren}
\end{equation}
So the NEC is
\begin{equation}
N^\mu N^\nu \langle 0| T_{\mu \nu}^{ren} |0 \rangle \geq 0
\label{scneccondition}
\end{equation}
and we would like to check if this condition is violated.

The left-hand side of Eq.~(\ref{scEeqnren}) is a tensor, hence
the right-hand side had better be a second rank tensor too. 
Since the only tensor available to us in de Sitter spacetime is
the metric, we immediately have:
\begin{equation}
\langle 0| T_{\mu \nu}^{ren} |0 \rangle = A g_{\mu\nu}
\label{Tvev}
\end{equation}
where $A$ is a constant. But then:
\begin{equation}
N^\mu N^\nu \langle 0| T_{\mu \nu}^{ren} |0 \rangle = 
    A g_{\mu \nu} N^\mu N^\nu = 0
\label{neceq0}
\end{equation}
and the NEC is not violated by a scalar field in de Sitter 
spacetime within the realm of semiclassical gravity.

\subsection{Fluctuations of the energy momentum tensor}
\label{fluctuations}

Fluctuations of the renormalized energy-momentum tensor 
may show NEC violations even though the expectation value
does not. Indeed, Eq.~(\ref{neceq0}) shows that the expectation
value of $N^\mu N^\nu T_{\mu\nu}^{ren}$ vanishes and hence
any fluctuations in it will violate the NEC. To calculate the 
magnitude of the fluctuations, we need to consider:
\begin{equation}
\langle 0| N^\mu N^\nu T_{\mu \nu}^{ren} (x) 
       N^\lambda N^\sigma T_{\lambda \sigma}^{ren} (x)|0 \rangle 
\nonumber
\end{equation}
However, since the operators are evaluated at the same spacetime
point, the fluctuations will be infinite. 

This infinity is not a crisis since, after all, we are interested 
in NEC violations not at a single spacetime point but in an entire
spacetime region. Hence we should consider fluctuations of a 
``smeared'' operator, which we can define as:
\begin{equation}
T_{W} \equiv 
  \int d^{4}y\sqrt{-g}\, W(y;x) N^\mu N^\nu T_{\mu\nu}^{ren}(y) 
\label{smearedoperator}
\end{equation}
where $W(y;x)$ is a window function of our choosing, centered at
$x$. Note that the smearing is both in space and time. 

The NEC is violated if
\begin{equation}
\sigma^2 \equiv \langle T_W^2 \rangle - \langle T_W \rangle^2 
    ~ > ~ \langle T_W \rangle^2 \equiv \mu^2
\label{twcondition}
\end{equation}

Even without doing any calculation, we can see that NEC violating 
fluctuations must occur. From Eq.~(\ref{neceq0}) we know that 
$\langle T_W \rangle =0$. The smearing in Eq.~(\ref{smearedoperator})
ensures that the operators in $\langle T_W^2 \rangle$ are multiplied 
at distinct spacetime points. The spacetime smearing ensures that
$\langle T_W^2 \rangle$ is finite. (To see this explicitly one 
needs to consider the integrals in more detail.) Also, we can
expect $\langle T_W^2 \rangle$ to be non-zero because there are 
no symmetries that force it to vanish. In fact, since the operator
$T_W$ is non-trivial, even if $\langle T_W^2 \rangle$ were to vanish
for some choice of window function, expectation values of yet higher 
powers of $T_W$, {\em e.g.} $\langle T_W^n \rangle$ for $n > 2$, could 
be considered. A non-vanishing result for any $n$ would indicate
NEC violating fluctuations.

In the case when the smearing scales are chosen to be given by the
horizon size $H^{-1}$, there is only one dimensionful quantity in 
the problem and dimensional arguments can be used. Hence, in this 
case,
\begin{equation}
\langle T_W^2 \rangle = \alpha H^8
\label{dimresult}
\end{equation}
where $\alpha \ne 0$ is a finite number. This result is already
interesting. Since the expectation of $T_W$ vanishes, it says
that there are both negative and positive fluctuations in $T_W$.
Assuming that either sign is equally likely, this implies that 
the NEC is violated with 50\% probability. The actual magnitude of 
the NEC violating fluctuation will depend on the parameter $\alpha$,
which will also depend on the smearing function. On spatial
and temporal scales given by the horizon, the magnitude of NEC
violation is $\sqrt{\alpha} H^4$.

A few new issues arise when we try to do better than the dimensional
estimate. First, for convenience, we choose the window function to be 
Gaussian in conformal time and space. 
\begin{equation}
W\left( \eta ,{\mathbf{r}}\right) =
         \frac{1}{\sqrt{-g}}\frac{a_{0}^{4}}{R^{3}\tau }
           W_{\eta }\left( \frac{\left| \eta -\eta _{0}\right| }
                        {a_{0}^{-1}\tau }\right) W_{r}
\left( \frac{\left|{\mathbf{r}}-{\mathbf{r}}_{0}\right| }{a_{0}^{-1}R}\right) 
\label{eq:Wdef}
\end{equation}
Here
\begin{equation}
a_{0}\equiv a\left( \eta _{0}\right) ,
\quad \tau \equiv H^{-1}\tanh (HT)
\end{equation}
and we have introduced the reference point
$(\eta _{0},{\mathbf{r}}_{0})$ with $\eta _{0}\neq 0$, 
since $\eta =0$ is a singular point of the conformal coordinate 
system. (In our coordinates 
$\sqrt{-g}=\left( H\eta \right) ^{-4}=a^{4}$.)
The normalization of the window and of the null vector 
\begin{equation} 
N^{\mu }= a_0 \left( H\eta \right) ^{2}\left[ 1,{\mathbf{n}} \right] .
\end{equation}
is chosen to make the final result independent of $\eta _{0}$. 
The time-dependent normalization of the null vector $N^{\mu }$ is 
such that $N^{\mu}$ is affinely parametrized. For simplicity, the 
three vector ${\mathbf{n}}$ will be chosen to be the unit radial 
vector. 

The parameter $\tau$ is defined so that averaging in conformal time 
$\eta $ with the window 
$a_{0}\tau ^{-1}W_{\eta }\left( a_{0}\tau ^{-1}|\eta -\eta _{0}|\right)$
corresponds to a window with proper time duration $T$ where the
relation between conformal time, $\eta$, and proper time, $t$, is:
\begin{equation}
\eta =-H^{-1}e^{-Ht}=-\left( Ha\left( t\right) \right) ^{-1}
\label{etaandt}
\end{equation}
Note that $H\tau <1$ always holds.
Likewise, the spatial window in Eq.~(\ref{eq:Wdef}) is such that in
the neighborhood of $\eta =\eta _{0}$ the proper length corresponding to
the spatial averaging is $R$.

In addition, we generalize the calculation to be more applicable
to inflationary cosmology where the de Sitter background changes
due to a slowly rolling scalar field. Then the scalar field is
given by:
\begin{equation}
\phi = \phi_0 + \delta \phi
\label{phi}
\end{equation}
where $\phi_0$ denotes the coherent state representing the
rolling field and $\delta \phi$ denotes quantum fluctuations.
Since the field value is now changing with time, 
$\mu \equiv \langle T_W \rangle \ne 0$ and is given 
by the kinetic energy in $\phi_0$:
\begin{equation}
\mu = \frac{1}{2} {\dot \phi}_0^2 \ .
\label{muequals}
\end{equation}

The calculation of $\sigma^2$ is now straightforward
albeit tedious, requiring clever estimation of certain integrals. 
The final result will be given only for $R=T=(\varepsilon H)^{-1}$. 
\begin{equation}
\sigma ^{2}\sim 
 \frac{
 H^{4}\dot{\phi }_{0}^{2}
 \max \left( c_{1}^{2}\varepsilon ^{2},c_{1}'^{2}\varepsilon ^{4}\right)
      } {\left( 2\pi \right) ^{2}}+
 \frac{c_{2}^{2}H^{8}\varepsilon ^{8}}{\left( 2\pi \right) ^{4}}
\label{sigma2}
\end{equation}
where $c_1$, $c_1'$ and $c_2$ are constants. We have evaluated these 
constants numerically for the Gaussian window function and find them 
to be of order unity. The last term in Eq.~(\ref{sigma2}) is new. It 
arises due to the expectation value of four creation and annihilation 
operators ($\langle a a a^\dag a^\dag \rangle$) and cannot be derived 
by simply considering root-mean-square fluctuations of the scalar 
field $\phi$. We can also compare the magnitude of the fluctuations 
to the square of the mean, as needed in Eq.~(\ref{twcondition}):
\begin{equation}
\frac{\sigma ^{2}}{\mu ^{2}}\sim 
 \frac{
 H^{4}\max \left( c_{1}^{2}\varepsilon ^{2},c_{1}^{'2}\varepsilon ^{4}\right) 
      }{\left( 2\pi \dot{\phi }_{0}\right) ^{2}}+
  \frac{c_{2}^{2}H^{8}\varepsilon ^{8}}
       {\left( 2\pi \dot{\phi }_{0}\right) ^{4}}.
\label{eq:sigma-ans}
\end{equation}

In the special case of an exactly de Sitter background
(${\dot \phi}=0$), Eq.~(\ref{sigma2}) gives:
\begin{equation}
\sigma ^{2}\sim 
 \frac{c_{2}^{2}H^{8}\varepsilon ^{8}}{\left( 2\pi \right) ^{4}}
\label{dSsigma2}
\end{equation}
Since $\mu =0$ in this case, the very fact that $\sigma$ is
non-vanishing implies that the NEC is violated.

It is fair to say that the detailed evaluation of $\sigma^2$
is not very crucial for us since here are only interested in
knowing whether NEC violations exist. This was evident from
Eq.~(\ref{dimresult}) itself. Yet the detailed calculation is
relevant when asking more quantitative questions. (What
is the typical magnitude of an upward jump that is coherent
on some given scale?) One subtlety in the detailed evaluation
is that it is easier to do the calculation when the window function 
is chosen to be a Gaussian in conformal time, but much harder 
if it is Gaussian in proper time. 

\section{Backreaction}
\label{backreaction}

Recall that our derivation in Sec.~\ref{nec} for the necessity 
of NEC violations in eternal inflation was based on the classical 
Einstein equations. The derivation could also be extended using 
the semiclassical equations provided we think of $\rho$ and $p$ 
in Sec.~\ref{nec} as being expectation values. However, what
we have shown above is that NEC violations only occur in
the fluctuations of the energy-momentum tensor
in de Sitter spacetime. Such fluctuations do not couple to
the metric by the classical or semiclassical Einstein equations. 
Then, what equations should we use to couple the metric and the
energy-momentum quantum operator? Is there a smeared version
of the Einstein or other equation? In other words, we need some
prescription to determine the backreaction of the quantum fluctuations 
on the metric.

There are potentially two new effects that can arise when
calculating the backreaction. First, we will need to let
the metric fluctuate as well. Without such metric fluctuations,
the semiclassical Einstein equations will hold and they imply 
that the Hubble expansion rate cannot grow as required in
eternal inflation. Once the metric is allowed to fluctuate,
there will be interactions with the scalar field fluctuations 
that will affect the NEC violating rate. (This is similar to
gravitational corrections occurring in instanton amplitudes.)
There will also be independent quantum fluctuations of the metric. 
For example, it is conceivable that even if one were to take a 
de Sitter background without any scalar field, the metric would 
fluctuate, and there would be regions that would expand slightly 
faster and others slightly slower. If so, one could presumably 
argue for an eternally inflating multiverse even without a scalar 
field. This would be quite interesting.
The second new effect is that the fluctuation amplitude falls 
off with length scale. So if there is an upward jump in
some local region, one might expect that the field outside
this region will jump down to compensate. This effect might
lead to interesting correlations of fluctuations. Another
way to state this new effect is that whatever the scalar field
energy-momentum tensor fluctuations may be, the semiclassical 
Einstein equations must still hold although in some average
sense. Hence there can be no ``overall'' eternal inflation and
local increases in the expansion rate must be accompanied by
other local decreases in the expansion rate.

The current treatment of the backreaction in the literature
assumes that if there is an NEC violating fluctuation, 
it simply resets the value of the scalar field on the potential
and the metric itself does not participate in the fluctuation.
After the fluctuation is over, we can revert to a classical 
description of the evolution with the reset value of the
scalar field. Then eternal inflation follows as described in 
Sec.~\ref{scenario}.

\section{Conclusions}
\label{conclusions}

The search for NEC violations in de Sitter spacetime was 
motivated by the classical Einstein equations. Within the 
realm of classical gravity and its semiclassical extension, 
no NEC violations were found. However, fluctuations of the 
energy-momentum tensor of a scalar field in de Sitter spacetime 
were found to violate the NEC. Hence one must look to the 
gravitational backreaction of energy-momentum tensor fluctuations 
to derive a convincing model of eternal inflation. In addition, 
one must also consider the quantum fluctuations of the metric 
itself. 

In the literature, backreaction calculations have been attempted
that fall into two categories. First is the backreaction of
classical fluctuations 
\cite{BriHar64,Isa68,HuxPazZha93,MukFelBra92}. 
In these calculations, the classical Einstein equations always 
hold but corrections to a background metric ({\em e.g.} inflationary
background) are evaluated in a perturbative manner. We have
already seen that classical physics cannot give the required
NEC violation and hence this approach cannot help with eternal 
inflation. The second category of backreaction calculations 
\cite{TsaWoo96,TsaWoo97} (summarized in \cite{Woo01})
considers the effects of quantum fluctuations on the spacetime,
and seems to be more suited to addressing eternal inflation.
Presently the pioneers of this approach find that the metric
fluctuations themselves -- in a de Sitter spacetime with only
a cosmological constant and without any scalar field -- have a 
very strong effect on the de Sitter background and can switch 
off the inflation or even lead to a period of deflation. They
understand this result physically in terms of the mutual
gravitational attraction of gravitons produced by the inflating 
background. With this physical picture, since these are quantum 
effects and hence probabilistic, there will be local fluctuations 
in the number of gravitons and hence in the local Hubble expansion
rate. One would have thought that the result of the calculation 
should then lead to eternal inflation even in the absence of a 
scalar field!

The eternal inflation scenario is seductive for a variety of 
reasons. At the moment, however, it rests on heuristic arguments.
More rigorous calculations are needed to make it quantitative
and convincing.

\begin{acknowledgments}
This work was supported by DOE.
\end{acknowledgments}

\end{document}